\documentclass[12pt]{iopart}
\usepackage{graphicx,epsfig}
\usepackage{xcolor}
\usepackage{bm}
\usepackage{cite}
\usepackage{hyperref}
\usepackage{caption}
\usepackage{subcaption}

\eqnobysec
\usepackage{dcolumn}

\begin{document}

\title{Inside the Jaynes-Cummings sum}

\author{S.I. Pavlik}
\ead{sipavlik@yahoo.com}
\address{Independent Researcher, Zaporizhzhia 69104, Ukraine}
\vspace{10pt}
\date{\today}

\begin{abstract}
It is shown that the atomic inversion in the Jaynes-Cummings model has an exact representation as an integral over the Hankel contour. For a field in a coherent state, the integral is evaluated using the saddle point method. The trajectories of saddle points as a function of time are on the branches of the multi-valued Lambert function. All of them start at the initial moment of time, but make the maximum contribution to the inversion at different times. If the collapse and the first revival are clearly distinguished, then subsequent revivals are determined by the comparable contributions of several trajectories.

\end{abstract}


\maketitle


\section{Introduction}
One of the simplest models of interacting systems in the framework of quantum electrodynamics consists of the two-level system with level separation $\omega_0$ interacting with a single-mode quantized electromagnetic field with a frequency $\omega$. In the rotating wave approximation, this model is described by the Jaynes-Cummings Hamiltonian \cite{jaynes1963comparison,Jay} 
\begin{equation}
	H=\omega \left(a^{+}a+\frac{1}{2}\right)+ \frac{1}{2}\omega_{0}\sigma_{3}+\lambda(\sigma_{+}a+\sigma_{-}a^{+}),\label {eq:E1}
\end{equation}
where $a$ and $a^{+}$ are the annihilation and creation operators of the photon field, $\lambda$  is the coupling constant. We use the Pauli matrices $\sigma_{i}$   ($i=1,2,3$ ) to describe the two-level atom and the notation $\sigma_{\pm}=1/2(\sigma_{1} \pm \sigma_{2})$. For simplicity we have assumed the coupling $\lambda$ to be real. Various aspects of the Jaynes-Cummings model are reviewed in numerous reviews, see for example \cite{ShoKni,larson2021jaynes}.
 
The Jaynes-Cummings Hamiltonian \eref{eq:E1} is easily diagonalizable \cite{Jay}, since it connects only a pair of basis states $| n \rangle |1\rangle$ and $| n-1 \rangle |2\rangle$, where $|1\rangle$  and $|2\rangle$  are the ground and exited state, and $|n\rangle$ is an eigenstate of the photon-number operator $a^{+}a$. Thus, for a certain value of $n>0$, the eigenfunctions will be a superposition of a pair of basic states \cite{Jay}, 
\begin{equation}\label {eq:E2}
	\eqalign{|+, n \rangle=\cos\alpha_{n}|2\rangle|n-1\rangle+\sin\alpha_{n}|1\rangle|n\rangle,\cr
	|-, n \rangle=-\sin\alpha_{n}|2\rangle|n-1\rangle+\cos\alpha_{n}|1\rangle|n\rangle,}		
\end{equation}
with
\begin{eqnarray}
	\cos\alpha_{n}=\frac{1}{\sqrt{2}}\sqrt{\left(1+\frac{\Delta}{2\Omega_{n}}\right)},\quad\sin\alpha_{n}=\frac{1}{\sqrt{2}}\sqrt{\left(1-\frac{\Delta}{2\Omega_{n}}\right)}.\nonumber
\end{eqnarray}
Here we define $\Delta=\omega_{0}-\omega$ as the detuning parameter, and $\Omega_{n}=\sqrt{(\Delta^{2}/4)+\lambda^{2}n}$ is the so called Rabi frequency. The eigenvalues of the Hamiltonian \eref{eq:E1} (for $n>0$) are
\begin{equation}
	H|\pm,n\rangle=\lambda_{\pm}(n)|\pm,n\rangle,\qquad\lambda_{\pm}(n)=\omega n\pm\Omega_{n}.
\end{equation}	
The eigenfunctions (\ref{eq:E2}) with the ground state, $H|1\rangle|0\rangle=-(\Delta/2)|1\rangle|0\rangle$, form a complete set of eigenfunctions of the Hamiltonian (\ref{eq:E1}).

Time evolution of an arbitrary initial state $|\Psi(0)\rangle$ is given by
\begin{eqnarray}
	\fl |\Psi(t)\rangle=\rme^{-\rmi Ht}|\Psi(0)\rangle=\rme^{\rmi\frac{\Delta}{2}t}\langle1|\langle 0|\Psi(0)\rangle|1\rangle|0\rangle +\sum_{n=1,k=\pm}^\infty \rme^{-\rmi\lambda_k(n)t}\langle k,n|\Psi(0)\rangle |k,n\rangle. \label {eq:E3}	
\end{eqnarray}	
Using Eq.~(\ref{eq:E3}), one can get the temporal behavior of any dynamic variables. We will be interested in atomic population inversion $\langle\sigma_{3}(t)\rangle=\langle\Psi(t)|\sigma_{3}|\Psi(t)\rangle$. If we assume that the atom is initially in the ground state $|1\rangle$, and the field is in a state $\sum_{n=0}^\infty\gamma_{n}|n\rangle$, where $W_{n}=|\gamma_{n}|^2$ determines the distribution of photons, the atomic inversion is given by the Jaynes-Cummings sum:
\begin{equation}
	\langle\sigma_{3}(t)\rangle=-\sum_{n=0}^\infty W_{n}\frac{\mu+n\cos(2\sqrt{\mu+n}\lambda t)}{\mu+n},\label {eq:E5}
\end{equation}
where the notation, $\mu=\Delta^2/4\lambda^2$, is used.
This is the exact solution. It would seem that it remains only to plot graphs under various conditions. But it turned out that this sum hides manifestations of interesting phenomena at different time scales. The magic of these phenomena is hidden into the the Jaynes-Cummings sum.

If the field is initially prepared in a coherent state, that is, 
\begin{equation}
	W_{n}=\frac{|\alpha|^{2n}}{n!}\exp(-|\alpha|^{2})\label {eq:E5a}
\end{equation}
is the Poisson distribution, $|\alpha|^{2}$ is the average photon number, then the initial value of the atomic inversion quickly decays, it has been called Cummings collapse \cite{Cum}, but it reappears many times at a later time \cite{EbNaSa,NarSaEb}. The envelope of the Rabi oscillations periodically revives and collapses. These phenomena occur in the real world and has been observed in experiments \cite{PhysRevLett.58.353,PhysRevLett.76.1800,PhysRevLett.76.1796}. 
 
To get a qualitative description of this phenomenon, the following simple arguments can be used  \cite{scully1997quantum}. The system starts from a correlated initial state, then the phases of different components diverge from each other, which leads to collapse. At this stage, the determining factor is the variance of the photon distribution, $ \Delta  n^{2}=\langle n^{2}\rangle-\langle n\rangle^{2} $, where $\langle n\rangle=\sum_{n=0}^\infty nW_{n}$ is the average photon number. The collapse time $t_{0}$ can be obtained qualitatively from the condition that the phases of the components over the distribution width during this time diverge by a value on the order of unity, $(\Omega_{\langle n\rangle+\Delta n}-\Omega_{\langle n\rangle-\Delta n})t_{0}=1$. Thus, the assumption is valid that the distribution near its average value takes on a maximum value and the relative width $\Delta n/\langle n\rangle$ decreases with an increase in the average number of photons. In this case, we get a finite collapse time. Subsequent revivals occur when the components of the sum that are near the maximum of the distribution are in phase, $(\Omega_{\langle n\rangle}-\Omega_{\langle n\rangle-1})t_{k}=2\pi k$, where $k=1,2...$ denotes a revival number. For a coherent state, the conditions for the occurrence of collapse and revival are satisfied, since the mean and the variance are equal, so $\Delta n=\sqrt{\langle n \rangle}$, and the relative width is $\Delta n/\langle n\rangle=1/\sqrt{\langle n\rangle}$. 
Surprisingly, these simple estimates for the time scales of collapse and revival are consistent with the results of a more rigorous approach \cite{NarSaEb}.

In order to get a more rigorous justification, one can start by analyzing the sum in Eq.~(\ref{eq:E5}), then applying a suitable summation method. But at the same time, certain difficulties arise due to the fact that the summation index is under the square root. Therefore, it is advisable to represent the sum as an integral and only then apply approximate methods to evaluate the resulting integral. The integral representation of the sum in Eq.~(\ref{eq:E5}) is determined by the discrete photon distribution $W_{n}$. Some discrete distributions can be approximated by continuous distributions. In fact, when $\langle n \rangle \sim 10$, a Gauss distribution is a good approximation of Poisson distribution. Then, converting the sum to an integral, a suitable method for approximating the integral can be used. In \cite{NarSaEb}, using the Euler-Maclaurin formula, without estimating the remainder term, the summation over $n$ is replaced by one integral,  which are then evaluated by the saddle point method. Thus, three approximations must be made along this path. First, when moving from a discrete to a continuous distribution, then when replacing the sum with an integral, and finally, when we calculate the integral. Nevertheless, excellent, classic results are obtained. In \cite{FleSchl}, the infinite sum changed an infinite sum of integrals with the help of the Poisson summation formula. Each revival  is connected with one term of this infinite sum integrals, which are then calculated by the stationary phase method. In \cite{KLIMOV1999100}, for the initial thermal field, using the Abel-Plana formula, the Jaynes-Cummings sum is replaced as a sum of two integrals, where the first integral represents the initial collapse and the second integral gives quantum corrections. For the coherent initial field, a similar problem was considered in \cite{Azu}. In \cite{KarKar}, to study the behavior of the collapse and revival of Rabi oscillations, the authors proposed an analytical treatment based on the application of some number theoretic results to the approximation of trigonometric sums. Our aim in this paper is to find an exact integral representation for the sum and only then use approximate methods to calculate the resulting integral.
 
The article is organized as follows. In Sec.~\ref{sec:jc}, with the help of integral representation for the Bessel function of half-integer order, the Jaynes-Cummings sum (\ref{eq:E5}) is replaced by a contour integral in the complex plane. Moreover, in the resonant case, $\Delta=0$, we do not use any approximations at this step. In Sec.~\ref{sec:res} for a field in a coherent state, the resulting integral, for the case $\langle n\rangle\gg 1$, is evaluated using the saddle point method. The saddle points are found from an equation whose solution is the Lambert $W$ function. Since the Lambert function is a multi-valued function, a single-valued trajectory of a saddle point as a function of time is located on each branch of this function. Analyzing the trajectories of the saddle point as a function of time, we find the moments of time in the vicinity of which they make the maximum contribution to the atomic inversion. Since all saddle points are born at the initial moment of time, the contribution from them overlaps over time, which leads to the suppression of separate revivals. Using the Lambert function allows us to obtain expressions for collapse and revivals in a simple and straightforward way, making only one approximation. In Sec.~\ref{sec:nores}, we analyze the Jaynes-Cummings sum in the non-resonant case. Here already the transition from the sum to the integral required an additional approximation. We give expressions for the collapse and revival functions. Sec.~\ref{sec:summary} is devoted to a summary and an outlook.
\section{Integral representation of the Jaynes-Cummings sum
\label{sec:jc}}
Consider the resonance case, $\Delta=0$.  The atomic inversion is given by
\begin{equation}
	\langle\sigma_{3}(t)\rangle=-\sum_{n=0}^\infty W_{n}\cos(2\sqrt{n}\lambda t).\label {eq:E6}
\end{equation}
In what follows, we will use the replacing $\lambda t\rightarrow t$.
 
Since the summation index is under the square root, it is desirable to use such a representation of the trigonometric function so that the replacement occurs $\sqrt{n}\rightarrow n$. Taking into account the variety of integral representations of Bessel functions, we choose the equality $\cos{x}=\sqrt{\pi x/2}J_{-1/2}(x)$ \cite{WhiWa}, where $J_{-1/2}(x)$ is the Bessel functions of half-integer order. This is motivated by the fact that the Bessel functions have a suitable representation as a contour integral in which the argument is squared. Using the Schläfli contour integral for the Bessel function \cite{Cop,WhiWa}, we have
\begin{equation}
	\cos{x}=\frac{1}{2\sqrt{\pi}\rmi}\int_{-\infty}^{(0^+)} \frac{1}{\sqrt{z}}\exp\left({z-\frac{x^2}{4z}}\right) \rmd z, \label {eq:E7}
\end{equation}
where $\mid\arg z \mid<\pi$; the cut is along the negative real semi-axis. The path of integration is the Hankel contour coming from $-\infty$, turning upwards around $0$, and heading back towards $-\infty$ \cite{Cop}. The notation $(0^+)$ just means that the path goes round the origin in the positive sense. With Eq.~(\ref{eq:E7}), by inverting the order of integration and summation, the Jaynes-Cummings sum (\ref{eq:E6}) is rewritten as
\begin{equation}
	\langle\sigma_{3}(t)\rangle=-\frac{1}{2\sqrt{\pi}\rmi}
	\int_{-\infty}^{(0^+)}\frac{\rme^{z}}{\sqrt{z}}\sum_{n=0}^\infty W_{n} \rme^{-nt^{2}/z}\rmd z.\label {eq:E8}
\end{equation}

Note that at the initial moment of time $t=0$ we obtain an integral representation of the Gamma function (Hankel's representation \cite{Cop} ), so that the atomic inversion is equal to $-1$, as it should be. 
In the representation (\ref{eq:E8}) the sum is greatly simplified, since the generating function of the distribution of photons obtained under the integral sign is easily calculated for most probability distributions.
In addition, the convergence of the integral in Eq.~(\ref{eq:E8}) at the ends of the path is exponential. 
\section{Poisson photon distribution in the resonance case
\label{sec:res}}
If the field is initially in a coherent state (\ref{eq:E5a}), calculating the sum in Eq.~(\ref{eq:E8}), the atomic inversion is given by
\begin{equation}
\langle\sigma_{3}(t)\rangle=-\frac{|\alpha|\tau^{1/2}}{2\sqrt{\pi}\rmi}
\int_{-\infty}^{(0^+)}\frac{1}{\sqrt{z}}\rme^{|\alpha|^{2}\Phi(z)}\rmd z,\label {eq:E9}
\end{equation}
with
\begin{equation}
	\Phi(z)=\tau z+\rme^{-1/z}-1,\label {eq:E10}
\end{equation}
where $\tau=t^{2}/|\alpha|^{2}$ and $z\rightarrow zt^{2}$. Note that the origin is both a branch point and an essential singularity. 

For large $|\alpha|^{2}$, we use the saddle point approximation to the integral in Eq.~(\ref{eq:E9}). Numerous examples of asymptotic analysis of special functions based on the contour integral representation can be found in the book \cite{Olver}.

Saddle points are found from the equation
$\Phi^{'}(z_{0})=\tau +1/z_{0}^{2}\exp(-1/z_{0})=0$, where the prime denotes a derivative with respect to $z$. Solutions to this equation give paths, $z_{0}(\tau)$, in the complex plane. The contribution to the atomic inversion is 
\begin{equation}
	\langle\sigma_{3}(t)\rangle\cong \rmi\sqrt{\frac{\tau}{2|\Phi^{''}(z_{0})|z_{0}}} \rme^{|\alpha|^2 \Phi(z_{0})+\rmi\theta}, \label {eq:E11}
\end{equation}
where the angle is determined from the condition $2\theta+arg\Phi^{''}(z_{0})=\pi+2\pi n$, which includes the second derivative. So far, we have used such variables to emphasize the connection with the integral representation of the Bessel functions. Now it's time to make a change of variables. If we introduce a new variable $F=-1/(2 z_{0})$, then
\begin{equation}
	\phi(F)=\Phi\left(-\frac{1}{2F}\right)=-\frac{\tau}{2F}+ \rme^{2F}-1 \label {eq:E12}
\end{equation}
and
\begin{equation}
	\Phi^{''}=-4F\tau\left(1+F\right).\label {eq:E13}
\end{equation}
Here $F$ is a solution of the equation
\begin{equation}
	\frac{\tau}{4}=-F^{2}\rme^{2F},\label {eq:E14}
\end{equation}
which is obtained from $\Phi^{'}=0$. Using Eq.~(\ref{eq:E12}) and Eq.~(\ref{eq:E13}) in Eq.~(\ref{eq:E11}), we get
\begin{equation}
	\langle\sigma_{3}(t)\rangle\cong\frac{\rmi}{2}\sqrt{\frac{-F}{|F(1+F)|}} \rme^{|\alpha|^2 \phi(F)+\rmi\theta} + c.c., \label {eq:E15}
\end{equation} 
where '$ c.c.$' denotes the complex conjugate, this follows from the fact that the atomic inversion is a real function.
The angle $\theta$ is defined as $\theta=\pi/2 -\arg(-F)/2-\arg(1+F)/2$, where we have used Eq.~(\ref{eq:E13}).
Thus, the basic expression for the atomic inversion associated with a certain saddle point trajectory $F$ is given by
\begin{equation}
	\langle\sigma_{3}(t)\rangle\cong -\frac{1}{\sqrt{|(1+F)|}} e^{|\alpha|^2 \Re\phi(F)}\cos\left(|\alpha|^2\Im\phi(F)-\frac{1}{2}\arg(1+F)\right), \label {eq:E16}
\end{equation}
where the functions $\Re\phi(F)$ and $\Im\phi(F)$ are the real and imaginary parts of $\phi(F)$. This is a universal expression that describes both collapse and revival. Now we need to find solutions of Eq.~(\ref{eq:E14}) at different times and use it in Eq.~(\ref{eq:E16}).
\subsection{Lambert W-function: the key to collapse and revivals
\label{sec:resa}}
Removing the square on the right side, Eq.~(\ref{eq:E14}) takes the general form
\begin{equation}
	u=w\rme^{w}, \label {eq:E17}
\end{equation}
where $u=x+\rmi y$ and $w=\xi+\rmi\eta$. The solution of this equation, $w=W\left(u\right)$, is the Lambert W-function \cite{CorLa}. The trajectory of the saddle point $w= F\left(\tau\right)$ is obtained from the solution of Eq.~(\ref{eq:E17}), assuming that $x=0$ and $y=\pm\left(\tau^{1/2}/2\right)$, where the plus-minus signs define two families of trajectories in the complex plane $w$. Note that in what follows we will define the solution $F\left(\tau\right)$ for $y=+\left(\tau^{1/2}/2\right)$, and the complex conjugate $F^{*}\left(\tau\right)$ for $y=-\left(\tau^{1/2}/2\right)$. Since the equation~(\ref{eq:E17}) is the square root of Eq.~(\ref{eq:E14}), we will consider two instances of the $w$-plane, $F\left(\tau\right)$ is located on one copy, and the complex conjugate $F^{*}\left(\tau\right)$ lives on the other. In other words, the desired trajectories, $F\left(\tau\right)$ and $F^{*}\left(\tau\right)$, are the images of the positive and negative imaginary $u$-axis in the $w$-plane, respectively. Since the Lambert W-function is a multi-valued function, each branch contains a single trajectory, either $F\left(\tau\right)$ or $F^{*}\left(\tau\right)$.

In order to cut the $w$-plane into branches of the $W$ function, we find the images of the negative real $u$-axis in the $w$-plane. This follows from the analogy with the logarithm \cite{CorLa}. From Eq.~(\ref{eq:E17}) it is easy to find curves in the $\left(\xi,\eta\right)$- plane that satisfy the conditions $x < 0$ and $y=0$. Note that the equation $y=0$ has two solutions, one of which is $\eta=0$. Thus, it is necessary to cut along the $\xi$-axis from $-\infty$ to the intersection with the cut at the point $w=-1$ (the solution of $u_{w}=0$), which is the image of the second-order branch point $u=-1/\rme$.
\begin{figure}[ht]
	\centering\includegraphics[width=\linewidth]{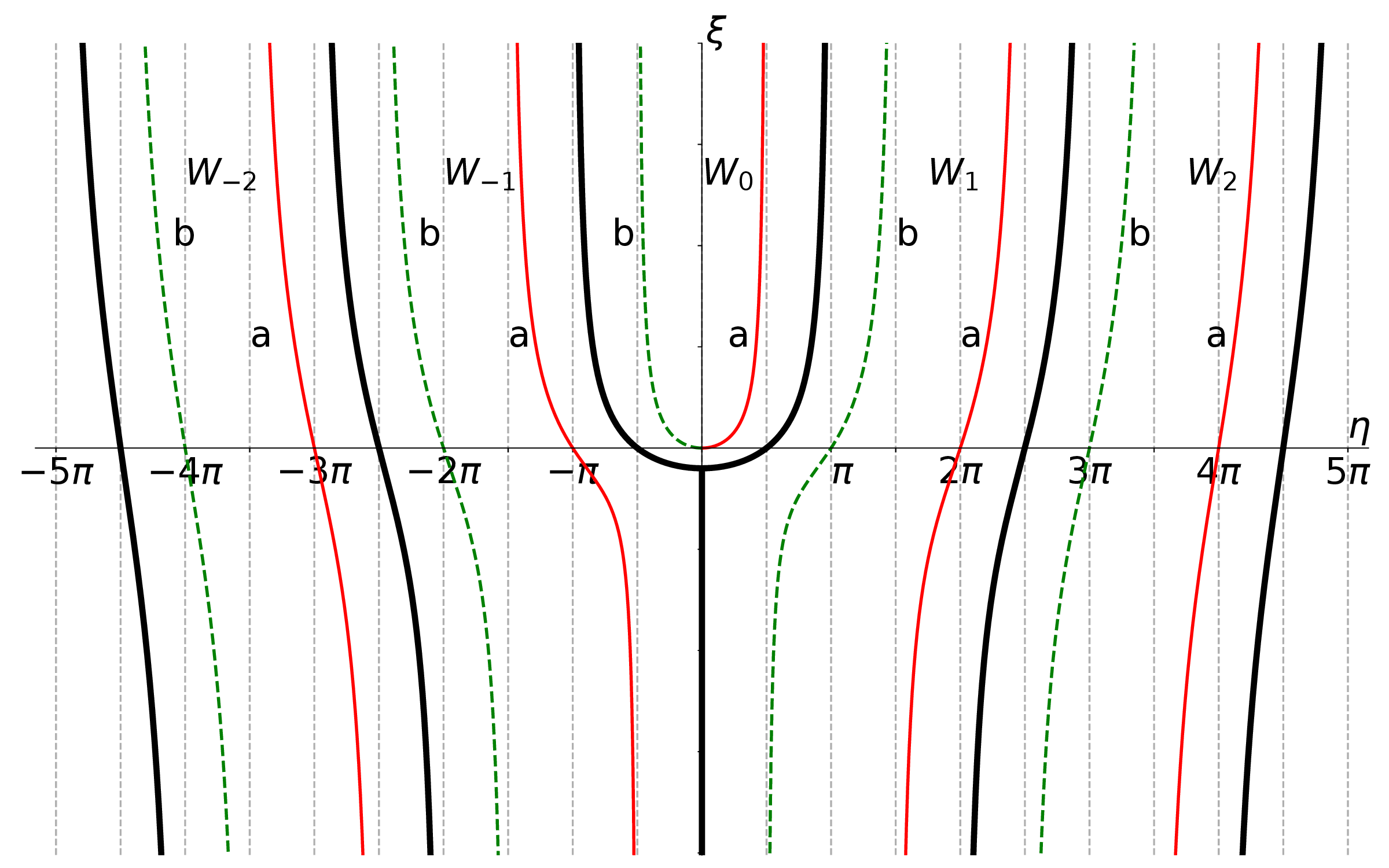}
	\caption{Trajectories of saddle points on the branches of the Lambert W-function. The function $F\left(\tau\right)$ (solid lines) on one instance of the complex plane (a), and $F^{*}\left(\tau\right)$ (dashed lines) on the other (b). Heavy solid lines are branch cuts. The branches $W_{k}$ are numbered $k=0,\pm 1,...$. The arrow of time points upwards.}
	\label{fig:Lambert}
\end{figure}
\begin{figure}[ht]
	\centering\includegraphics[width=\linewidth]{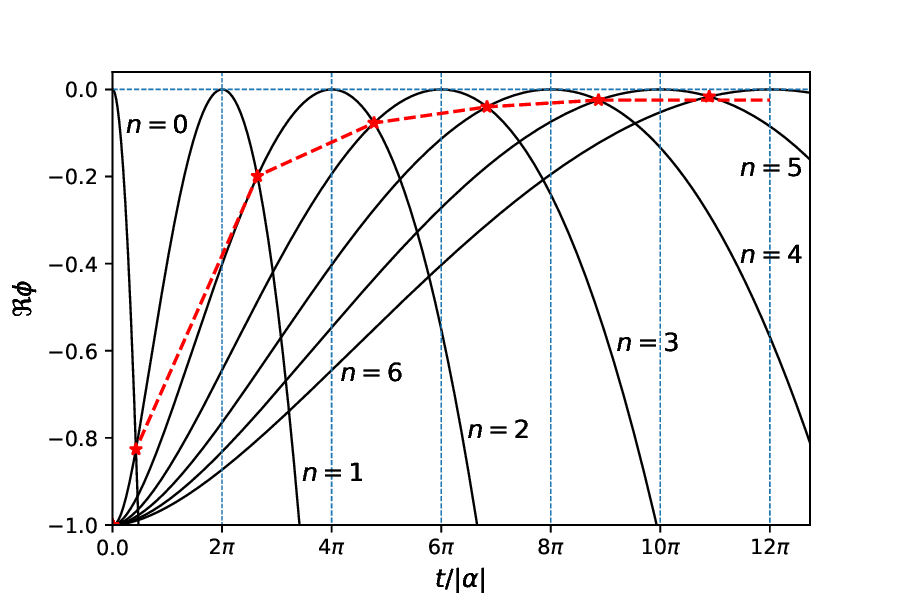}
	\caption{Contributions (solid lines) from the trajectories of saddle points to the amplitude function $\Re\phi$ from Eq.~(\ref{eq:E12}); $n=0$ corresponds to collapse, $n>0$ - to revivals. The curve through the intersection points of neighboring "bursts" is shown as a dashed line.}
	\label{fig:saddle}
\end{figure}
Figure~\ref{fig:Lambert} shows the branches of the multivalued Lambert W-function. The plane is cut into branches on which $W\left(u\right)$ is an analytic function. We separately consider two instances (branches) of the complex plane. On one branch, we consider the trajectory of the saddle point $F\left(\tau\right)$, which is a solution of Eq.~(\ref{eq:E17}) with $u=+i\tau^{1/2}/2$, and on the other, we consider its complex conjugate partner. The trajectories intersect the $\eta$-axis at points $\pi n$, $n=\pm 1,\pm 2, \dots$, in the $\omega$-plane that correspondent $\tau^{1/2}=2\pi n, n>0$. All saddle points are born at $\tau=0$ and move along the trajectory from bottom to top. On the principal branch $W_{0}$,  the starting point is at $w=0$, on the others at $\xi\rightarrow -\infty$.
\begin{figure}[ht]
	\centering\includegraphics[width=\linewidth]{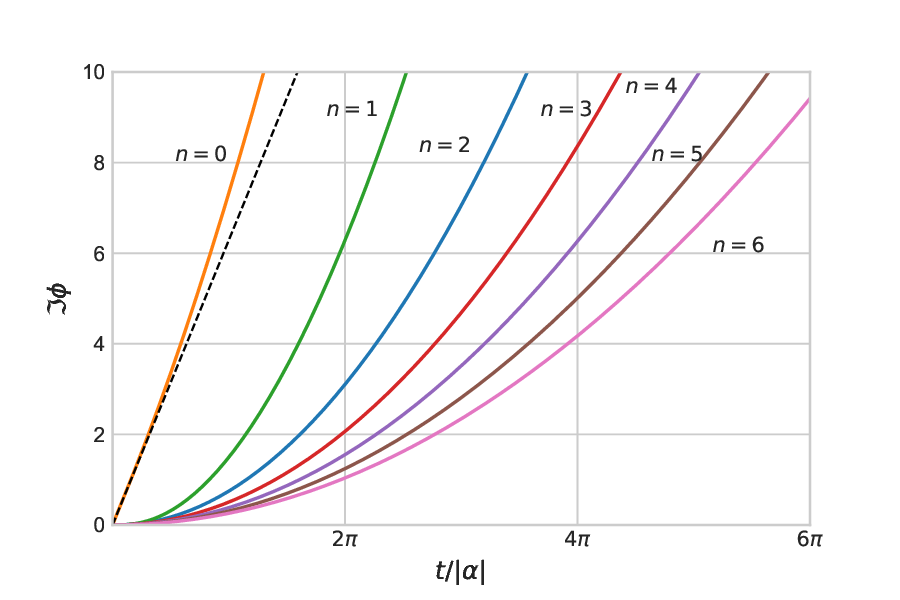}
	\caption{Contributions (solid lines) from the trajectories of saddle points to the amplitude function $\Im\phi$ from Eq.~(\ref{eq:E12}); $n=0$ corresponds to collapse, $n>0$ - to revivals. Linear function $2t/|\alpha|$ is shown as a dashed line.}
	\label{fig:image}
\end{figure}

\begin{figure}
	\centering
	\begin{subfigure}[a]{0.9\textwidth}
		\includegraphics[width=\textwidth]{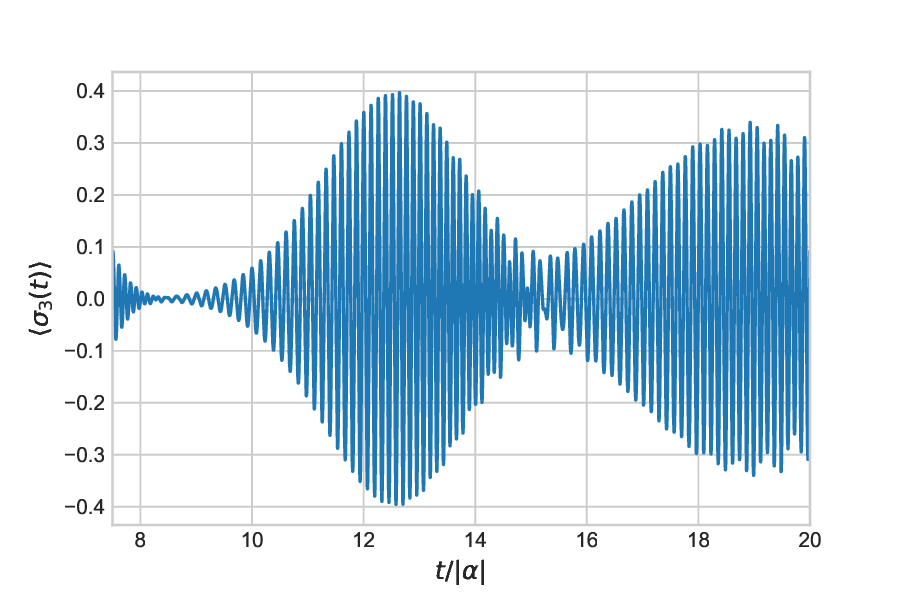}
		\caption{}
		\label{fig:ex_a}
	\end{subfigure}
	\begin{subfigure}[b]{0.9\textwidth}
		\includegraphics[width=\textwidth]{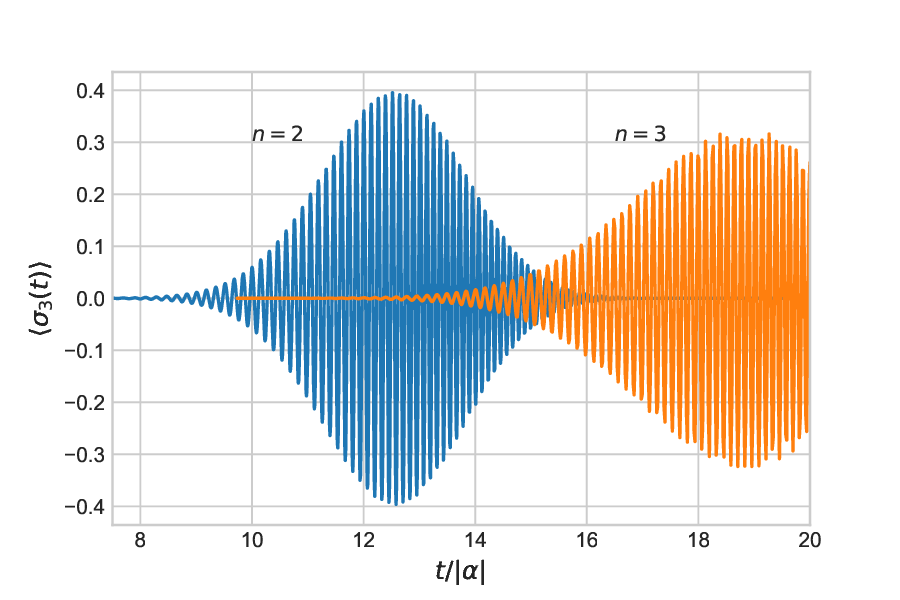}
		\caption{}
		\label{fig:ap_b}
	\end{subfigure}
	\caption{The second and third revivals for a coherent state as a function of the scaled time $t/|\alpha|$ for $|\alpha|=5$. The atomic inversion $\langle\sigma_{3}\rangle$ is plotted on the vertical axes. The exact sum (\ref{eq:E6}) shown in (a) is in good agreement with the approximate analytical result based on Eq.~(\ref{eq:E16}) and depicted in (b).}	
	\label{fig:exap}
\end{figure}

It is clear that the asymptotic properties of the atomic inversion, Eq.~(\ref{eq:E16}), is determined only by the behaviour of $\Re \phi$ near the times when this function takes the maximum value. Further, it is easy to see from Eq.~(\ref{eq:E12}) that $\Re\phi=0$ is the maximum value that is taken at $w_{n}=i\pi n$ ($n=\pm 1,\pm 2, \dots$) and $t_{n}=2\pi n |\alpha|, n>0$.
From Figure~\ref{fig:saddle}, we can see that the functions $\Re\phi$, depending on the time, start near $t=0$, but reach a maximum at different times. Thus, their width grows. If for $n=0,1$ one can speak of an isolated time-separated contribution to the atomic inversion from the trajectories of saddle points, then the subsequent curves overlap with each other. The intersection points of the curves are located at $t\approx4\pi |\alpha|n(n+1)/(2n+1)$ ($n\geq 1$), their value is getting closer and closer to the maximum value of the function $\Re\phi$. In this case, one cannot speak of separate revivals, since neighboring bursts largely overlap with each other. As $|\alpha|$ grows, the scale changes, but not the overall picture.

Figure~\ref{fig:image} shows the imaginary part $\Im\phi$ as a function of the scaled time $t/|\alpha|$ at different branches of the Lambert function. On the principal branch ($n=0$), the linear function $2t/|\alpha|$2  is a good approximation to $\Im\phi$, at least in short time. From the figure~\ref{fig:saddle} it can be seen that this corresponds to collapse. On the remaining branches $n\neq 0 $ the quadratic function $t^2/2\pi n |\alpha|^2 $ is a good approximation to $\Im\phi$ for all times. These curves correspond to revivals.

Thus, by looking at the curves in Figure~\ref{fig:saddle} and using Eq.~(\ref{eq:E16}), one can describe the behavior of atomic inversion. So, the population at the initial moment of time is equal to $-1$, which corresponds to $\Re\phi=0$, but quickly goes out. We are observing a collapse. Then the inversion flashes, but does not go out completely. However, the first revival is clearly marked. Then the inversion lights up again, but with a smaller amplitude due to the pre-exponential factor in Eq.~(\ref{eq:E16}), and barely noticeably blinks with a certain period. Thus, there will be no brightly distinguished revivals, except for the initial moments of time. 

To confirm this qualitative picture, in Figure~\ref{fig:exap} we plot the atomic inversion as a function of dimensionless time, both the exact sum Eq.~(\ref{eq:E6}) in Figure~\ref{fig:ex_a}, and the approximate analytical result (\ref{eq:E16}) in Figure~\ref{fig:ap_b} for the second and third revivals. In Figure~\ref{fig:ex_a} on the left is a fragment from the first revival. As we suggested above, the first revival is clearly marked and the amplitude of the atomic inversion between the first and second revivals practically drops to zero. The transition between the second and third revivals (at $t\approx4\pi |\alpha|n(n+1)/(2n+1)=15.08$, for $n=2$) in the first and second figure agrees well with each other. On the other hand, the contributions from each saddle trajectory to the amplitude function $\Re\phi$ near the maxima at times $t/|\alpha|=2\pi n$ are sharp enough that it is possible to expand the function $\phi$ to obtain approximate expressions for the atomic inversion.
\subsection{Collapse
\label{sec:collapse}}
Consider the trajectory of the saddle point $F(\tau)$ on the principal branch $W_{0}$. The function $F(\tau)$ is the solution of Eq.~(\ref{eq:E17}) with $u=\rmi\tau^{1/2}/2$, $F(\tau)=W(\rmi\tau^{1/2}/2)$. Since the right side of the Eq.~(\ref{eq:E17}) is regular in the neighborhood $w=0$, we can use the Lagrange's formula for the reversion of a power series \cite{Cop}, that is,  
\begin{equation}
	W(u)=\sum_{n=1}^{\infty}c_{n}u^{n},\label {eq:E18}
\end{equation}
where 
\begin{eqnarray}
	c_{n}=\frac{1}{n!}\frac{\rmd^{n-1}}{\rmd w^{n-1}}\left(\frac{w}{w\rme^{w}}\right)^{n}\bigg\vert_{w=0}=\frac{(-n)^{n-1}}{n!}.\nonumber
\end{eqnarray}
The radius of convergence of the series (\ref{eq:E18}) is equal to $1/\rme$. Recall that 
$u=-1/\rme$ is the second-order branch point.

For $\tau\ll1$, you can take only the first two terms in Eq.~(\ref{eq:E18}), $F\left(\tau\right)\approx \rmi\tau^{1/2}/2+\tau/4$. Substituting this expression
in Eq.~(\ref{eq:E12}) we immediately arrive at
\begin{equation}
	\phi\approx-\frac{\tau}{2}+2\rmi\tau^{1/2}.\label {eq:E19}
\end{equation}
Since here $\Im F\ll1$, then $arg(1+F)\approx 0$, Eq.~(\ref{eq:E16}) rewritten as
\begin{equation}
	\langle\sigma_{3}(t)\rangle\approx- \rme^{-t^2/2} \cos\left(2|\alpha|t\right). \label {eq:E20}
\end{equation}
The formula (\ref{eq:E20}) is a well-known expression for the Cummings collapse. Note that the cosine phase in Eq.~(\ref{eq:E20}) is good agreement with the linear approximation for $\Im \phi $ ($n=0$) shown as dashed line in Figure~\ref{fig:image}, near the initial time. 
\subsection{Revivals
\label{sec:revivals}}
Now consider the contribution to the atomic inversion (\ref{eq:E16}) from saddle points on other branches of the Lambert function, $W_{\pm k}$ with $k>0$. As we found out earlier, see also Figure~\ref{fig:saddle}, the function $\Re\phi=0$ takes its maximum value when $\tau_{n}=4\pi^2 n^2$, $F(\tau_{n})=F_{n} =\rmi\pi n$ ($n> 0$). Strictly speaking, we must take $F$ for odd $n> 0$ (see else on Figure~\ref{fig:saddle}), and $F^*$ for even $n$, but since a pair of complex conjugate saddle points contribute to the final result, we will use $F$ for any positive values $n$.

The Taylor series of a function in Eq.~(\ref{eq:E12}) about a point $\tau=\tau_{n}$ is given by
\begin{equation}
	\phi(\tau,F(\tau))\approx\phi(\tau_{n},F_{n})+ \frac{\rmd\phi}{\rmd\tau}\bigg\vert_{\tau_{n}}(\tau-\tau_n)+\frac{1}{2}\frac{\rmd^2\phi}{\rmd\tau^2}\bigg\vert_{\tau_{n}}(\tau-\tau_n)^2+\dots \label {eq:E21}
\end{equation} 
where the total derivatives are taken along the curves defined by the equation $\phi_{F}=0$, that is, 
\begin{eqnarray}
	\frac{\rmd\phi}{\rmd\tau}=\phi_{\tau},\;\;\frac{\rmd^2\phi}{\rmd\tau^2}=\phi_{\tau\tau}+\phi_{F\tau}F_{\tau},\nonumber
\end{eqnarray}
where $F_{\tau}=-\phi_{F\tau}/\phi_{FF}$; we used the rule of implicit differentiation. For $\tau$ near  $\tau_n$ in Eq.~(\ref{eq:E21}), one can replace $(\tau-\tau_n)^2\approx 4\tau_n(\tau^{1/2}-\tau_n^{1/2})^2$. Calculating the derivatives of $\phi$ in Eq.~(\ref{eq:E21}), substituting into Eq.~(\ref{eq:E16}), we obtain
\begin{eqnarray}
\fl \langle\sigma_{3}(t)\rangle\approx-\frac{1}{(1+\pi^2 n^2)^{1/4}} \exp{\left[-\frac{(t-t_n)^2}{2(1+\pi^2 n^2)}\right]} \cos\left[\frac{t^2}{2\pi n}-\frac{(t-t_n)^2}{2\pi n(1+\pi^2 n^2)}-\frac{\pi}{4}\right].\label {eq:E22}
\end{eqnarray}
Here we used the approximation $arg(1+F_n)\approx \pi/2$. It is clear that the second term under the cosine can be neglected, so that we get the familiar expression for the revivals \cite{EbNaSa,NarSaEb,FleSchl} (subject to $1+\pi^2 n^2\approx \pi^2 n^2$). The phase $t^2/2\pi n $ can also be checked by comparison with the versus time curves (for $n>0$) in Figure~\ref{fig:image}), and as we can see, excellent agreement is found.
\section{Poisson photon distribution in the non-resonant case
\label{sec:nores}}
In the resonance case, we used the exact expression for the Jaynes-Cummings sum based on the contour integral representation of the Bessel function. We then applied the saddle point method to approximate the resulting integral. In the non-resonant case, the integral representation (\ref{eq:E7}) is also used, but now there is no way to sum in Eq.~(\ref{eq:E5}), as in the transition from Eq.~(\ref{eq:E8}) to Eq.~(\ref{eq:E9}). At this stage, one more approximation is required. 
 \subsection{Approximate integral representation of the atomic inversion}
Consider the Jaynes-Cummings sum (\ref{eq:E5}) with a field in a coherent state. Let's split the sum into two parts. For the time-independent part, it is easy to obtain the following expression:
\begin{equation}
	-\rme^{-|\alpha|^{2}}\sum_{n=0}^\infty\frac{|\alpha|^{2n}}{n!}\frac{\mu}{\mu+n}=-1 +\frac{\rme^{-|\alpha|^{2}}}{|\alpha|^{2\mu}}\int_{0}^{|\alpha|^{2}}x^{\mu}\rme^{x}\rmd x,\label {eq:E23}
\end{equation}
where we used the formula
\begin{equation}
	\sum_{n=0}^\infty\frac{a^n}{n!}\frac{1}{b+n}=\int_{0}^{1}x^{b-1}\rme^{ax}\rmd x \label {eq:E24}
\end{equation} 
and then integrated by parts.  As $|\alpha|^{2}\gg 1$, using the leading asymptotic behavior \cite{bender1999advanced} of the integral in Eq.~(\ref{eq:E23}), it is easy to see that the time-independent part of the Jaynes-Cummings sum is negligible compared to the second part in Eq.~(\ref{eq:E5}). However, the next term in the asymptotic expansion of the integral on the right side of in Eq.~(\ref{eq:E23}) is $\mu/|\alpha|^2$. Naturally, this procedure works for small $\nu$. Thus, applying the integral representation (\ref{eq:E7}), the time-independent part of the atomic inversion is written as follows:
\begin{equation}
	\langle\sigma_{3}(t)\rangle=-\frac{1}{2\sqrt{\pi}\rmi}
	\int_{-\infty}^{(0^+)}\frac{\rme^{z-|\alpha|^{2}}}{\sqrt{z}}\sum_{n=0}^\infty \frac{|\alpha|^{2(n+1)}}{n!}\frac{\rme^{-t^{2}(\mu+1+n)/z}}{\mu+1+n}\rmd z.\label {eq:E25}
\end{equation}
Note that by differentiating Eq.~(\ref{eq:E25}) with respect to $t^2$, one can obtain an exact expression for the sum in last expression. But on this way there are difficulties with integrating the resulting expression with respect to $t^2$. Thus, we proceed directly to the summation in Eq.~(\ref{eq:E25}). Using Eq.~(\ref{eq:E24}) with $z\rightarrow zt^{2}$, we have 
\begin{equation}
	\langle\sigma_{3}(t)\rangle=-\frac{t}{2\sqrt{\pi}\rmi|\alpha|^{2\mu}}\int_{-\infty}^{(0^+)}\rmd z\frac{1}{\sqrt{z}}\rme^{zt^2-|\alpha|^{2}}\int_{0}^{|\alpha|^{2}\rme^{-1/z}}\rmd x x^{\mu}\rme^{x}.\label {eq:E26}
\end{equation}
Using the leading asymptotic behavior of the last integral in Eq.~(\ref{eq:E26}) (for $|\alpha|^{2}\gg 1$), the approximate atomic inversion (\ref{eq:E5}) is given by
\begin{equation}
	\langle\sigma_{3}(t)\rangle\approx -\frac{|\alpha|\tau^{1/2}}{2\sqrt{\pi}\rmi}
	\int_{-\infty}^{(0^+)}\frac{1}{\sqrt{z}}\rme^{|\alpha|^{2}\Phi(z,\nu)}\rmd z,\label {eq:E27}
\end{equation}
with
\begin{equation}
	\Phi(z,\nu)=\tau z-\frac{\nu}{z}+\rme^{-1/z}-1,\label {eq:E28}
\end{equation}
where $\nu=\mu/|\alpha|^2$. Note that Eq.~(\ref{eq:E27}) is an exact integral representation for atomic inversion (\ref{eq:E9}) at resonance ($\nu=0$).

Applying the saddle point method, we obtain
\begin{equation}
	\langle\sigma_{3}(t)\rangle\approx-\frac{1}{\sqrt{|f_{\nu}|}} \rme^{|\alpha|^2 \Re\phi_{\nu}}\cos\left(|\alpha|^2\Im\phi_{\nu}-\frac{1}{2}\arg f_{\nu}\right),\label {eq:E29}
\end{equation}
where
\begin{equation}
	\phi_{\nu}=-\frac{\tau}{2F_{\nu}}+2F_{\nu}\nu+\rme^{2F_{\nu}}-1 \label {eq:E30}
\end{equation}
and
\begin{equation}
	f_{\nu}=1+F_{\nu}+\frac{4F^{3}_{\nu}\nu}{\tau}.\label {eq:E31}
\end{equation}
Trajectories of saddle points $F_{\nu}(\tau)$ are determined from the equation $\Phi^{'}(z_{\nu})=0$ ($z_{\nu}=-1/2F_{\nu}$), which takes the form
\begin{equation}
	\frac{\tau}{4}=-F^{2}_{\nu}\left(\nu+\rme^{2F_{\nu}}\right),\label {eq:E32}
\end{equation}
As in Sec.~\ref{sec:resa}, we need to analyze the solutions of this equation in order to get a picture of collapse and revival in the non-resonant case.
\subsection{Off-resonant continuation of the Lambert function
\label{sec:offLam}}
As follows from Eq.~(\ref{eq:E32}), a generalization of Eq.~(\ref{eq:E17}) to the non-resonant case is given by
\begin{equation}
	u=w(\nu+\rme^{2w})^{1/2}. \label {eq:E33}
\end{equation}
As in the resonant case, there exists a function $w=W(u,\nu)$ that satisfies Eq.~(\ref{eq:E33}). The branches $W_{k}$ are defined for $|u|\rightarrow\infty$ by Eq.~(\ref{eq:E17}) and $W_{k}(u,\nu)\rightarrow\ln u+2\pi \rmi k$. In other words, far from the imaginary $\eta$-axis in the $w$-plane, the cuts look like in Figure~\ref{fig:Lambert}. The intersection points of the cuts with the $\eta$-axis do not shift compared to the resonant case, but the behavior of the cut curves themselves will be slightly modified under the influence of $\nu$. Recall that the boundaries between the branches, when $k\neq 0$, are images of the negative $u$-axis in the $w$-plane. For the principal branch, the cut is $\left(-\infty, u_{0}\right]$, where $u_{0}$ is obtained from Eq.~(\ref{eq:E33}) at $w_{0}$, which is determined from the equation $u_{w}|_{w_{0}}=0$, that is, 
\begin{equation}
	\nu+\rme^{2w_{0}}(1+w_{0})=0. \label {eq:E34}
\end{equation} 
After changing the variables, it is easy to convert Eq.~(\ref{eq:E34}) to the form (\ref{eq:E17}), and then apply the Lagrange formula \cite{Cop} (see Eq.~(\ref{eq:E18})), as a result we get
\begin{equation}
	w_{0}=-1-\frac{1}{2}\sum_{n=1}^{\infty}\frac{(n)^{n-1}}{n!}(2\rme^{2}\nu)^{n}.\label {eq:E35}
\end{equation}
As the parameter $\nu$ increases, $w_{0}$ shifts from minus one down along the negative $\xi$-axis (see Figure~\ref{fig:Lambert}). When the dimensionless detuning $\nu$ reaches a critical value $\nu_{0}=1/2\rme^{3}$, the derivative singularity of $w_{0}$ appears and the second-order branch point falls down. This is just an interesting fact, since the trajectory of the saddle point on the principal branch is not affected in any way. However, in the non-resonant case, the trajectories of the saddle points change significantly. This is because, due to the square root in the equation~(\ref{eq:E33}), the branch points $1/2\ln\nu+\rmi\pi\left(n+1/2\right)$ ($ n$ integer) appear on the branches of $W(u,\nu)$. Note that, as $\nu\rightarrow 0$, these branch points go to  $\xi\rightarrow -\infty$.

As in Sec.~\ref{sec:res}, we assume that $F_{\nu}(\tau)=W(\rmi\tau^{1/2}/2,\nu)$. On the principal branch, the trajectory of the saddle point, $F_{\nu}(\tau)$, starts at $w=0$, $\tau=0$. On the other branches, the trajectories start near the branch points, then the curves intersect the $\eta$-axis points $\pi n$, $n=\pm 1,\pm 2, \dots$, in the $\omega$-plane that corresponding to $\tau_{n}^{1/2}=2\pi n(1+\nu)^{1/2}, n> 0$. The function $\Re \phi_{\nu}$ takes on a maximum value at $t_{n}=2\pi n |\alpha| (1+\nu)^{1/2}$ that corresponds to the revivals. 
\subsection{Collapse}
Consider $F_{\nu}(\tau)$ on the principal branch $W_{0}$. Applying the Lagrange inversion theorem to the mapping (\ref{eq:E33}), we get
\begin{equation}
	F_{\nu}(\tau)=\sum_{n=1}^{\infty}c_{n}(\nu)\left(\rmi\frac{\tau^{1/2}}{2}\right )^{n},\label {eq:E36}
\end{equation}
where
\begin{equation}
	c_{n}(\nu)=\frac{1}{n!}\frac{\mathrm{d}^{n-1}}{\mathrm{d}w^{n-1}}\left(\nu+e^{2w}\right)^{-n/2}\bigg\vert_{w=0}.\label {eq:E37}
\end{equation}
As in Sec.~\ref{sec:collapse}, for $\tau\ll1$, you can take only the first two terms in Eq.~(\ref{eq:E36}), $F\left(\tau\right)\approx \rmi\tau^{1/2}/2(1+\nu)^{1/2}+\tau/4(1+\nu)^{2}$. Inserting this into Eq.~(\ref{eq:E30}), we have
\begin{equation}
	\phi_{\nu}\approx-\frac{\tau}{2(1+\nu)}+2\rmi\tau^{1/2}(1+\nu)^{1/2}.\label {eq:E38}
\end{equation}
In a given approximation, Eq.~(\ref{eq:E31}) is $f_{\nu}\approx 1$ and $\arg f_{\nu}\approx 0$.

Putting it all together into Eq.~(\ref{eq:E29}), we get the inversion near the initial time
\begin{equation}
	\langle\sigma_{3}(t)\rangle\approx- \rme^{-t^2/2(1+\nu)} \cos\left(2|\alpha|t\right(1+\nu)^{1/2}), \label {eq:E39}
\end{equation}
that describes the collapse.
\subsection{Revivals}
As we found out in Sec.~\ref{sec:offLam}, the trajectories of the saddle point intersect the $\eta$-axis points $\pi n$ ($n=\pm 1,\pm 2, \dots$) that correspondent $t_{n}=2\pi n |\alpha| (1+\nu)^{1/2}$ ($n>0$). However, a significant difference from the resonance case is related to the starting point of the saddle point trajectory on the $w$-plane, which is now near the branch points $w_{n}=1/2\ln\nu+\rmi\pi\left(n +1/2\right)$. For $\nu\rightarrow 0$ the overall picture shown in Figure~\ref{fig:saddle} is restored. 

In order to consider the saddle points on the branches $W_{k} ($k$\neq0$), we use the recipes of Sec.~\ref{sec:revivals}. Expanding the function $\phi_{\nu}$, Eq.~(\ref{eq:E30}), near $t_n$, taking into account Eq.~(\ref{eq:E29}), we obtain
\begin{equation}
	|\alpha|^{2}\Re\phi_{\nu}=-\frac{1}{2}\frac{1+\nu}{\pi^{2}n^{2}+(1+\nu)^2}(t-t_n)^2,\label {eq:E40}	
\end{equation}
\begin{equation}
	|\alpha|^{2}\Im\phi_{\nu}=2\pi n\mu+\frac{t^2}{2\pi n}-\frac{1}{2\pi n}\frac{(1+\nu)^2}{\pi^{2}n^{2}+(1+\nu)^2}(t-t_n)^2 \label {eq:E41}
\end{equation}
and
\begin{equation}
	|f_{\nu}|=\left (1+\frac{\pi^2 n^2}{(1+\nu)^2} \right)^{1/2}.\label {eq:E42}
\end{equation}
It is easy to obtain from Eq.~(\ref{eq:E31}) that $\arg f_{\nu}\approx\pi/2$.
Near $t\approx t_n$, we can neglect the third term on the right side in Eq.~(\ref{eq:E41}).
Thus, the term $2\pi n\mu$  is the most significant factor signaling the features of the non-resonant case. 
\begin{figure}[ht]
	\centering\includegraphics[width=\linewidth]{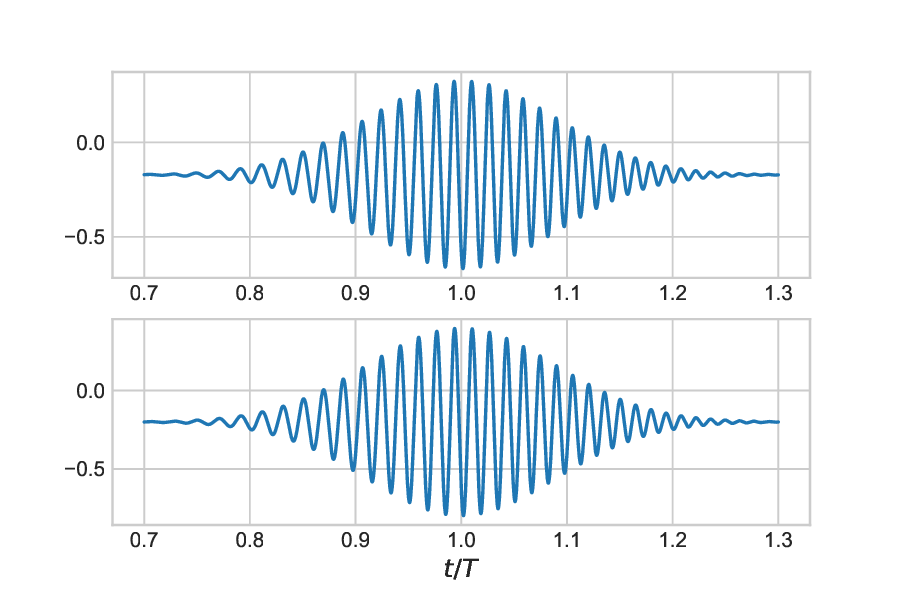}
	\caption{First quantum revival for a coherent state as a function of the scaled time $t/T$ ($T=2\pi |\alpha| (1+\nu)^{1/2}$) for $|\alpha|=5$ and $ \nu=0.2$ in the non-resonant case. The atomic inversion $\langle\sigma_{3}\rangle$ is plotted on the vertical axes. The top graph is the exact sum (\ref{eq:E5}), and the bottom graph is the approximate result (\ref{eq:E29}) using expressions (\ref{eq:E40}), (\ref{eq:E41}) and (\ref{eq:E42}). We added a time-independent part (\ref{eq:E23})   to Eq.~(\ref{eq:E29}); $-\nu$ is a good approximation for this.}
	\label{fig:fig5}
\end{figure}

Figure~\ref{fig:fig5} shows the atomic inversion $\langle\sigma_{3}\rangle$ as a function of scaled time $t/T$. In order to compare the exact and approximate solution, we consider the first quantum revival ($n=1$). As can be seen, the approximate analytical result based on Eq.~(\ref{eq:E29}) for $n=1$ (bottom graph) is in good agreement with the exact expression Eq.~(\ref{eq:E5}) (top graph). We must not forget that the integral representation of the inversion, Eq.~(\ref{eq:E27}), is itself an approximation, in contrast to the exact expression in the resonant case. 
\section{Discussions and summary
	\label{sec:summary}}
In the resonance case, using the integral representation for the Bessel function of half-integer order, the Jaynes-Cummings sum was written as a contour integral of the generating function for the photon distribution. Since the summation index is no longer under the square root, it is easy to calculate the sum for any reasonable distribution. For a field in a coherent state, an exact integral representation of the Jaynes-Cummings sum is obtained. As $|\alpha|^{2}\gg 1$, for the approximate calculation of the integral, the saddle point method was used. We have shown that the solution to the equation for the saddle point is the  multi-valued Lambert $W$-function. In other words, on each branch of this multi-valued function there is a single trajectory of a saddle point. The magic of collapse and revivals is hidden in the analytical structure of the Lambert function. The trajectory on the principal branch is associated with the collapse of the atomic inversion, on the other branches - with revivals. All saddle points start at the initial time, each on its own branch. At a certain point in time, it is necessary to choose the trajectory (with a complex conjugate) whose contribution to the integral has a maximum value. We have shown that the revivals are centered at certain times, at which the amplitude function of the atomic inversion reaches a single maximum on a certain trajectory. For collapse, the maximum is reached at the initial moment of time, and the atomic inversion quickly decays due to the exponential factor. For revivals, highs are reached at regular intervals. So, the point of intersection of neighboring trajectories is getting closer and closer to the maximum. The envelope of atomic inversion oscillations moves near the maximum, which decreases due to the exponential factor. Further, we have obtained, in the leading approximation, the asymptotics for the collapse and revivals near the maximum.

Unfortunately, in the nonresonant case, we were unable to obtain an exact integral representation for the Jaynes-Cummings sum. We needed one more explicit approximation. We derived an approximate integral representation in the form of a contour integral, for which we used the saddle point method. Compared to the resonance case, the overall picture has been preserved. The times of revivals have shifted due to the influence of the detuning parameter. An additional phase shift appeared in the phase.

An obvious and interesting continuation of this work would be to apply the formalism to other distributions of photons, to the calculation of correlation functions, and to other atomic properties (entropy, etc.).
\section*{Acknowledgements}
	I’d like to express gratitude to my family.

\end{document}